\documentclass[aps,pra,superscriptaddress,amsmath,amssymb,preprintnumbers,floatfix,showpacs,showkeys,12pt]{revtex4-1}
\usepackage{amssymb} \usepackage{epsfig}

\begin{document}
\title{A measure of tripartite entanglement in bosonic and fermionic
  systems } \author{Fabrizio~Buscemi }
\email{fabrizio.buscemi@unimore.it} \affiliation{Dipartimento di
  Elettronica, Informatica, e Sistemi, Universit\`{a} di Bologna,
  Viale Risorgimento 2, 40136 Bologna, Italy} \affiliation{ARCES, Alma
  Mater Studiorum, Universit\`{a} di Bologna, Via Toffano 2/2, 40125
  Bologna, Italy} \author{Paolo~Bordone} \affiliation{ Dipartimento di
  Fisica, Universit\`{a} di Modena e Reggio Emilia, 41125, Modena,
  Italy} \affiliation{Centro S3, CNR-Istituto di Nanoscienze, Via
  Campi 213/A, Modena 41125, Italy}

\begin{abstract}
  We describe an efficient theoretical criterion suitable for the
  evaluation of the tripartite entanglement of any mixed three-boson
  or -fermion state, based on the notion of the entanglement of
  particles for bipartite systems of identical particles. Our approach
  allows one to quantify the accessible amount of quantum correlations
  in the systems without any violation of the local particle number
  superselection rule.  A generalization of the tripartite negativity
  is here applied to some correlated systems including the
  continuous-time quantum walks of identical particles (both for
  bosons and fermions) and compared with other criteria recently
  proposed in the literature. Our results show the dependence of the
  entanglement dynamics upon the quantum statistics: the bosonic
  bunching results into a low amount of quantum correlations while
  Fermi-Dirac statistics allows for higher values of the entanglement.

\end{abstract}

\maketitle

\section{Introduction}

The notion of the entanglement in bosonic or fermionic systems has
been the subject of many recent discussions and controversy since it
rises some conceptual questions about the nature of the quantum
correlations appearing in such systems\cite{Schie, Gittings, Zana,
  Wise,Sasaki}. The main difficulty appearing in the definition of a
criterion apt to classify and quantify the amount of quantum
correlations is due to the intrinsic indistinguishability of the
particles supposed to be entangled, which results in the
symmetrization or antisymmetrization of the wavefunction. This
stimulated great interest in the scientific community and lead to
various proposals of entanglement criteria for identical particles.
In particular, a number of works present in the literature have been
focused on the analysis of bipartite entanglement~\cite{Schie,Zana,
  Wise,Dow}. In the paper by Schliemann \emph{et al.}~\cite{Schie},
the amount of quantum correlations between two fermions is evaluated
by considering the fermionic analogous of the Schmidt rank, namely the
Slater rank, which is given by the number of Slater determinants
needed to expand the entangled states.  The Schliemman's criterion has
been used to evaluate the entanglement in a number of bipartite
systems of physical interest~\cite{bus2,bus3, Lamata}, still it is
discussed since it does not seem to behave correctly under one-site
(local) and two-site (non local) transformations~\cite{Gittings}. In
the approach developed by Zanardi~\cite{Zana}, the entanglement is
evaluated in terms of the quantum correlations between modes by
mapping the Fock space of the modes themselves into qubit states.  It
has been shown that such a criterion overestimates the entanglement
between the two parties of the quantum system since it conflicts with
the local number particle superselection rule~\cite{Wise}. In order
not to violate the latter, Wiseman and Vaccaro~\cite{Wise} have
proposed an operational criterion for bipartite entanglement of
identical particles. They define as entanglement of particles the
maximum amount of quantum correlations which can be extracted by means
of local operation on the modes from the two parties of the system and
then set in standard quantum registers (composed of distinguishable
qubits)~\cite{Wise,Dow}. Only recently a coherent picture on the
entanglement of indistinguishable particles somehow including all the
above criteria has been presented in the literature~\cite{Sasaki}.

Though the entanglement of multipartite systems suggests more
possibilities and phenomena with respect to the bipartite case, so far
it has not been widely investigated. Its quantification would
undoubtedly represents a key ingredient to understand a number of
physical phenomena involving correlated many-particle systems such as
quantum phase transitions, quantum Hall effect, high temperature
superconductivity~\cite{Oste,Hui,Vedra}.  Among the various physical
systems of identical particles, particular attention has been devoted
to the analysis of multipartite entanglement in a non-interacting
fermion gas.  It has been shown that the multiparticle entanglement
can be built only out of two-fermion entanglement~\cite{Lunkes}. The
existence and the nature of the entanglement in systems of
noninteracting three fermions have also been analyzed by means of
different parameterized entanglement
witnesses~\cite{Levai,Vertesi,Habiban} that enable to detect two
different classes of tripartite entangled states, namely
Greenberger-Horne-Zeilinger (GHZ) and $W$ states.  By using the
Zanardi approach~\cite{Zana}, that is by mapping the Fock space of a
systems of fermions into the isomorphic mode state, a geometric
measure of entanglement for $N$ fermions with spin $\frac{1}{2}$ has
also been proposed~\cite{Lari}. A general entanglement criterion
applicable to both bosonic and fermionic systems has recently been
introduced though it is defined only with reference to the measurement
setup~\cite{Sasaki}.

In this work, we discuss an entanglement criterion suitable both for
three-fermion and for three-boson mixed states. We follow the basic
concepts of the method proposed by Wiseman and Vaccaro~\cite{Wise} for
the analysis of the bipartite entanglement in systems of identical
particles.  Our criterion does not violate the particle local number
superselection rule.  In fact here the entanglement between three
subsets, say $A$, $B$, $C$ is given by the average, over the local
particle number, of the amount of quantum correlations existing among
the standard quantum registers of three subsets each with a definite
local particle number. Given the use of concept of standard quantum
registers, the tripartite negativity (TPN), recently introduced to
quantify the tripartite entanglement of non identical
subsystems~\cite{Sabin,Anza}, is here taken as a measure of the
entanglement exhibited by three-boson or three-fermion mixed states.
In order to get a better understanding of the criterion, first we
compare and contrast it to other approaches, evaluating the quantum
correlations in many-particle systems. Then we apply it to investigate
the entanglement dynamics in a simple model mimicking the
continuous-time quantum walks (QWs)~\cite{Kon} of three identical
particles (bosons or fermions) in an one-dimensional structure.
Specifically, QWs of indistinguishable elements constitute the ideal
systems to test and validate our theoretical approach since in them
the exchange terms lead to the formation of correlations between the
particles even in absence of interaction, as shown by the experimental
investigations on two-photon propagation in lattices of coupled
waveguides~\cite{Bromberg,Peruzzo}.

The paper is organized as follows: In Sec.~\ref{Theory} we define the
tripartite entanglement criterion in boson and fermion systems with
reference to other entanglement measures in systems of identical
particles.  The main properties of the TPN as a measure of the amount
of quantum correlations in a six-mode system and its comparison with
the geometric measure introduced in Ref.~\cite{Lari} are discussed in
Sec.~\ref{trin}.  In Sec.~\ref{Phys} we evaluate the time evolution of
the TPN in QWs of three particles propagating in a lattice for two
different cases: particles following the Bose-Einstein statistics and
particles following the Fermi-Dirac statistics.  Finally, we present
our conclusions in Sec.~\ref{Conclu}.

\section{Tripartite entanglement criterion}\label{Theory}
In this section, we introduce an entanglement criterion for boson and
fermion systems relying on the concept of entanglement of
particles~\cite{Wise}.

In order to represent the mixed quantum state of $N$ identical
particles, we adopt the occupation-number representation.  Thus, in a
system with $L$ modes an arbitrary mixed state of $N$-bosons
(fermions) can be written as~\cite{Jacoboni}:
\begin{equation}
  \rho= \sum_{\{n\},\{n^{\prime}\}} 
  f\big(\{n\},\{n^{\prime}\}\big) 
  |\{n\}\rangle \langle \{n^{\prime}\}|
\end{equation}
with the integers $n_i$ ($n_i^{\prime}$) of the set $\{n\}=n_1,\ldots,
n_i,\ldots, n_L$ $\big(\{n^{\prime}\}=n^{\prime}_1,\ldots,
n^{\prime}_i,\ldots, n^{\prime}_L\big)$ satisfying $n_1\ldots+
n_i\ldots+ n_L=N$ $\big(n^{\prime}_1\ldots+ n^{\prime}_i\ldots+
n^{\prime}_L=N\big)$.  The ket $|\{n\}\rangle$ is the state of the
Fock space with $n_i$'s particles in the single-particle modes $i$'s
and $f\big(\{n\},\{n^{\prime}\}\big)$ are the coefficients in the
superposition. While for bosons $n_i$'s range from 0 to $N$, for
fermions the occupation numbers are restricted to 0 and 1 by the Pauli
exclusion principle.

A formal equivalence between the Fock space and the space obtained as
tensor product of $L$-dimensional single-particles subspaces can be
established~\cite{Zana}.  In such a procedure, the occupation number
of each mode represents a distinct state of the mode itself.  Zanardi,
Lari~\emph{et al.}~\cite{Zana, Lari} defined as entanglement of a
system of identical particles the so-called entanglement of modes
($E_M$), namely the amount of quantum correlations appearing among the
occupation numbers of the modes.

As argued by some authors~\cite{Green,Green2}, an application of the
$E_M$, as it is, to the quantum states of many-body systems could lead
to misleading results since such a measure quantifies the entanglement
between modes and not between the particles.  In order to clarify this
point, let us consider a single particle in an equal superposition of
three modes, so that in the occupation-number representation its state
$|\chi \rangle$ reads
\begin{equation} \label{chi} |\chi \rangle=\frac{1}{\sqrt{3}}\left(
    |001\rangle +|010\rangle +|100\rangle \right).
\end{equation}
By using the geometric measure of the entanglement relying on the
isomorphism between the Fock space and the mode space~\cite{Lari}, we
can evaluate the tripartite entanglement $\epsilon_G$ between the
modes as:
\begin{equation} \label{eg} \epsilon_G=||\tau||-||\tau||_{\text{sep}},
\end{equation}
where
\begin{equation}
  ||\tau||=\sqrt{\sum_{i,j,k}^3 \left|\langle \chi| \sigma_i\otimes \sigma_j \otimes \sigma_k| \chi \rangle
    \right|^2},\qquad \text{and}\qquad ||\tau||_{\text{sep}}=1
\end{equation}
with $\{\sigma_i\}$ the set of three generators of the SU(2) group,
namely the Pauli operators, each acting on a single-mode space. After
straightforward calculations, one finds that the tripartite
entanglement of $|\chi \rangle$ is different from 0, specifically
$\epsilon_G$=$\sqrt{33}/3-1$. Such a result would seem to indicate a
single-particle non-locality stemming from the entanglement with the
vacuum~\cite{Green,Green2}. However, the above entanglement is not
apparent when the particle wavefunction $\chi(x)$ is considered in the
space configuration
\begin{equation} \chi (x)=\frac{1}{\sqrt{3}}\left( \psi_1(x) +
    \psi_2(x) +\psi_3(x) \right)
\end{equation}
with $\psi_i$ indicating the single-particle states.

To explain the emerging paradox, the appearance of non-locality has
been related to multi-particle effects~\cite{Green2}. Anyway, this
behavior arises some questions on the use of the criteria relying on
the $E_M$, since they do not always capture the true entanglement
between the parties of the system. As argued by Wiseman and
Vaccaro~\cite{Wise}, in particular such criteria fail to take into
account the local particle-number (LPN) superselection rule. In fact
each party of the system must be able to perform arbitrary local
operations on its modes in order to fully use the entanglement between
the modes.  Unless each subsystem possesses a definite number of
particles, such local operations violate the superselection rule for
the LPN and are not possible in practice.  Thus in general, the
entanglement measures relying on $E_M$ overestimate the available
entanglement.

A tripartite entanglement criterion obeying the local particle-number
superselection rule can be obtained by extending the operational
definition of bipartite entanglement of identical particles introduced
by Wiseman and Vaccaro~\cite{Wise,Dow}. Being Alice, Bob, and Charlie
the three parties of a quantum system in the mixed state $\rho$ each
accessing to a given set of modes, here we assume that any subsystem
possess a standard quantum register, namely a set of distinguishable
qubits in addition to the indistinguishable particles described by
$\rho$.  We define as tripartite entanglement $\epsilon_T$ the maximum
amount of quantum correlations that Alice, Bob, and Charlie can
produce between their standard quantum registers by means of local
operations. As a consequence of the LPN superselection rule, this
tripartite entanglement in place of quantum correlations between the
modes that Alice, Bob, and Charlie have access to, is given by
\begin{equation} \label{enta}
  \epsilon_T=\sum_{n_A,n_B,n_C}^{n_A+n_B+n_C=N}
  P_{n_A,n_B,n_C}\epsilon_{ABC}(\rho_{n_A,n_B,n_C}),
\end{equation}
where $\rho_{n_A,n_B,n_C}$ =$\Pi_{n_A,n_B,n_C}\rho\Pi_{n_A,n_B,n_C}$
is obtained from $\rho$ by means of the projectors $\Pi_{n_A,n_B,n_C}$
onto fixed LPN's states ($n_A$ for Alice, $n_B$ for Bob, and $n_C$ for
Charlie).
$P_{n_A,n_B,n_C}=\textrm{Tr}\left(\rho_{n_A,n_B,n_C}\right)$ is the
probability of finding $n_A$,$n_B$, and $n_C$ as a result of a
measurement of the local particle number by Alice, Bob, and Charlie,
respectively, while $\epsilon_{ABC}$ is an entanglement standard
measure which quantifies the degree of tripartite entanglement among
the three sets of modes each controlled by a party of the system. It
is worth noting that here is used a standard measure of the tripartite
entanglement, that is a measure of the tripartite entanglement of non
identical subsystems, since the standard quantum registers of Alice,
Bob and Charlie consist of distinguishable qubits.  Specifically,
since some good measures $\epsilon_{AB}$ of bipartite entanglement can
successfully be extended to multi-partite systems by considering
bipartite partitions of them~\cite{Yu}, we take as standard tripartite
entanglement measure the geometric mean of the entanglement measures
of the bipartitions of the system:
\begin{equation}\label{bip}
  \epsilon_{ABC}=\sqrt[3]{\epsilon_{A-BC}\epsilon_{B-AC}\epsilon_{C-AB}}
\end{equation}
in agreement with what proposed in the
literature~\cite{Love,Sabin,Anza}.  Thus, expression~(\ref{enta}) now
reads
\begin{equation} \label{enta2}
  \epsilon_T=\sum_{n_A,n_B,n_C}^{n_A+n_B+n_C=N}
  P_{n_A,n_B,n_C}\sqrt[3]{\epsilon_{A-BC}(\rho_{n_A,n_B,n_C})\epsilon_{B-AC}(\rho_{n_A,n_B,n_C})\epsilon_{C-AB}(\rho_{n_A,n_B,n_C})}.
\end{equation}

The tripartite entanglement $\epsilon_T$ does not violate the LPN
superselection rule.  In fact when Alice, Bob, and Charlie measure
their local particle numbers $n_A$,$n_B$, and $n_C$, the mixed state
$\rho$ collapses, with a given probability $P_{n_A,n_B,n_C}$, in
$\rho_{n_A,n_B,n_C}$. Local operations can be performed on the latter
without any restrictions in order to transfer its entanglement
$\epsilon (\rho_{n_A,n_B,n_C})$ to the standard quantum registers of
each subsystem.  Since $\epsilon_T$ is the weighted sum of the terms
$\sqrt[3]{\epsilon_{A-BC}(\rho_{n_A,n_B,n_C})\epsilon_{B-AC}(\rho_{n_A,n_B,n_C})\epsilon_{C-AB}(\rho_{n_A,n_B,n_C})}$,
the entanglement is not affected, on the average, by measurements on
the local particle number.

Unlike $\epsilon_G$, the tripartite entanglement of the single-photon
state written in Eq.~(\ref{chi}) according to our criterion is 0. By
considering the partition of the system in three subsystems each
controlling a single-mode state, the state $|\chi\rangle$ is given by
the linear superposition of three quantum states having different
LPN's and therefore the contribution to $\epsilon_T$ from each terms
is zero.  This result again supports the interpretation of
$\epsilon_T$ as a measure of the tripartite entanglement not among the
modes but among the particles of the systems, in agreement with the
analogous investigations performed for the bipartite
case~\cite{Wise,Dow}.

\section{TPN in a six-mode system}\label{trin}

In order to obtain $\epsilon_T$ according to the criterion given in
Eq.~(\ref{enta2}), the evaluation of the bipartite standard
entanglement $\epsilon_{I-JK}(\rho_{n_A,n_B,n_C})$ for all
bipartitions $I-JK$ of each possible set of LPN's is needed (with
$I$=$A,B,C$ and $JK$=$BC,AC,AB$, respectively).  Unfortunately, the
latter is a very challenging task in high-dimensional systems for
mixed state of multi-particle systems.

In this section, we intend to give a non ambiguous measure of the
tripartite entanglement $\epsilon_T$ which can be used in a practical
way in a simple system and then compare it with the geometric
measure~\cite{Lari}. To this aim, we restrict our analysis to the case
of three bosons (fermions) each in a six-mode single-particle space
$h_6$. The Fock space $H_6(3)$ of the full system is the totally
symmetric (antisymmetric) subspace of the tensor product $h_6^{\otimes
  3}$. By partioning the system in three subsets, each acceding two
modes, we note that only a few quantum states belonging to the Fock
space $H_6(3)$ give a non vanishing contribution to the tripartite
entanglement according to the criterion given in Eq.~(\ref{enta2}).
Let us examine the case of the set of mixed states $F_{(n_A=0)}:
=\text{span}\{ \rho_{(0,n_B,n_C)}\rangle\}$ ($n_B,n_C=0,1,2,3$ with
$n_B+n_C=3$) where the local number of particles possessed by Alice,
is zero.  Any state of $F_{(n_A=0)}$ can be written as ${|0 0\rangle
  \langle 0 0 |}_A \otimes \rho_{BC} $ and therefore it is
\emph{biseparable}, that is, it can be factorized in a term describing
Alice with no particle and in a term describing the Bob and Charlie
subsets with three particles. The bipartite entanglement relative to
the bipartition $A-BC$ is zero and, in turns, $\epsilon_T$ is
vanishing according to Eq.~(\ref{enta2}) . This implies that the
entanglement due to sets of quantum states with a LPN equal to zero is
vanishing.  On the other hand, the tripartite amount of quantum
correlations can be different from zero only when each party has one
particle, that is only for the quantum states belonging to the set
$F_{(n_a=1,n_b=1,n_c=1)}: =\text{span}\{ \rho_{(1,1,1)}\}$.  Since
each partition has two modes, the subspace spanned by the vectors of
$F_{(n_a=1,n_b=1,n_c=1)}$ turns out to be isomorphous to the
three-qubit Hilbert space
$\mathbb{C}^2\otimes\mathbb{C}^2\otimes\mathbb{C}^2$.  Thus the
entanglement of three identical particles (both bosons and fermions)
in this case can be evaluated in terms of the quantum correlations
between three qubits.

Different approaches have been developed to characterize the
tripartite entanglement in system of distinguishable qubits. The
three-tangle or residual entanglement related to squares of the
bipartite concurrences has been proposed~\cite{Coff}, even if its use
is questioned since it does not properly quantifies the three-party
entanglement for $W$ states~\cite{Jung}.  Alternatively, as stated in
the previous section, standard tripartite entanglement can be
estimated by means of some good measures of the amount of quantum
correlations in bipartite systems whose Hilbert space is
$\mathbb{C}^2\otimes\mathbb{C}^4 $.  It should be noticed that in this
case both von Neumann entropy and concurrence presents
weaknesses~\cite{Sabin}. The former is an appropriate measure only for
pure states, the latter, even if it is well defined for mixture states
of two qubits, is applicable to higher dimensions only for pure
states. A valid measure of tripartite entanglement for non pure-state
can be obtained by using the negativity. From the Eq.~(\ref{bip}), the
tripartite negativity (TPN) $\mathcal{N}_{ABC}$ can be defined as
\begin{equation}\label{bip2}
  \mathcal{N}_{ABC}=\sqrt[3]{\mathcal{N}_{A-BC}\mathcal{N}_{B-AC}\mathcal{N}_{C-AB}},
\end{equation}
where the bipartite negativities are given by
${N}_{I-JK}=\sum_i|\gamma_i(\rho^{T_I})|-1$ with
$\gamma_i(\rho^{T_I})$ the eigenvalues of $\rho^{T_I}$ which is the
partial transpose related to the subsystem $I$ of the total density
matrix.

By inserting the TPN of Eq.~(\ref{bip2}) in Eq.~(\ref{enta2}), the
tripartite entanglement of a mixed state $\rho$ of three bosons
(fermions) in a six-mode system, can be easily evaluated from
\begin{equation} \label{enta3} \epsilon_T=
  P_{1,1,1}\sqrt[3]{\mathcal{N}_{A-BC}(\rho_{1,1,1})\mathcal{N}_{B-AC}(\rho_{1,1,1})\mathcal{N}_{C-AB}(\rho_{1,1,1})},
\end{equation}
once is known the projection of $\rho$ on the subspace where the local
number of each party is fixed to 1, namely
$\rho_{1,1,1}$=$\Pi_{1,1,1}\rho\Pi_{1,1,1}$.

In order to illustrate some features of the entanglement measure
introduced in Eq.~(\ref{enta3}) in comparison with the geometric
measure $\epsilon_G$, here we consider a six-mode three-fermion system
in the state
\begin{equation} \label{phi} |\Phi\rangle=\cos\alpha \cos\beta
  |010101\rangle + \cos\alpha \sin\beta |101010\rangle+
  \frac{\sin\alpha}{\sqrt{2}} |111000\rangle +
  \frac{\sin\alpha}{\sqrt{2}} |000111\rangle
\end{equation}
where $\alpha,\beta$ are phases ranging from 0 to $\pi$.  If we
consider the following partition of the system $A$=$\{1,2\}$,
$B$=$\{3,4\}$, and $C$=$\{5,6\}$, we note that $|\Phi\rangle$ is a
linear superposition of states, each with a given LPN for any set of
modes.  When $\alpha$=0,$\pi$ and $\beta$=$(\pi/4),(3\pi/4)$, only the
first two terms are non vanishing and a GHZ state with each party of
the system containing a particle, namely, $n_A$=$n_B$=$n_C$=1, is
obtained. On the other hand for $\alpha$=$(\pi/2)$, a specific LPN
cannot be ascribed to the subsystems.  Our goal is to evaluate the
amount of quantum correlations of $|\Phi\rangle$ according to
$\epsilon_T$ and $\epsilon_G$. The latter can be calculated again from
the Eq.~(\ref{eg}), where now~\cite{Lari}:
\begin{equation}
  ||\tau||=\sqrt{\sum_{i,j,k=1}^{15} 8\left|\langle \Phi| \lambda_i\otimes \lambda_j \otimes \lambda_k| \Phi \rangle
    \right|^2},\qquad \text{and}\qquad ||\tau||_{\text{sep}}=6\sqrt{6},
\end{equation} with  $\{\lambda_i\}$ indicating  the set of fifteen  generators of the SU(4) group~\cite{Fuh}
acting on the two  sites of one  subset. 
\begin{figure}[h]
  \begin{center}
    \includegraphics*[width=\linewidth]{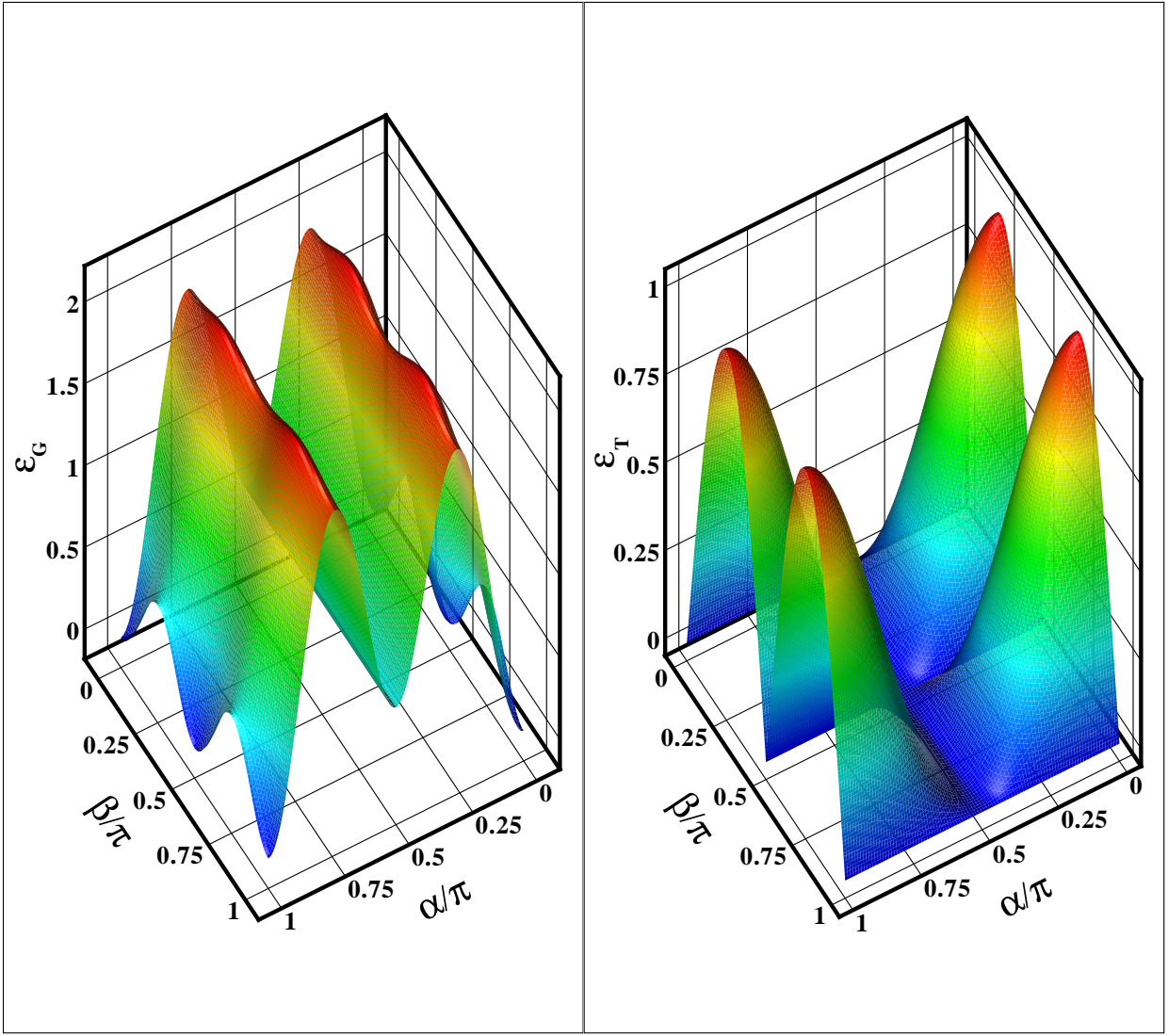}
    \caption{\label{fig1} (Color online) The tripartite entanglement
      $\epsilon_G$ (left panel) and $\epsilon_T$ (right panel) as a
      function of the phases $\alpha$, and $\beta$ ranging from 0 to
      $\pi$. Specifically, the quantum correlations exhibited by the
      three-fermion state $|\Phi\rangle$ of Eq.~(\ref{phi}) among the
      three parties, $A$=$\{1,2\}$, $B$=$\{3,4\}$, and $C$=$\{5,6\}$
      are here considered.}
  \end{center}
\end{figure}

Fig.~\ref{fig1} displays $\epsilon_T$ and $\epsilon_G$ as a function
of $\alpha$ and $\beta$. The two entanglement measures show a
different behavior and in agreement with the theoretical predictions
of the previous section the geometric measure is always greater than
the TPN. The latter is vanishing for any value of $\beta$ when
$\alpha=\pi/2$; in this case, $|\Phi\rangle$ becomes the linear
superposition of two terms, each corresponding to different LPN's thus
yielding zero entanglement.  At $\alpha$=0,$\pi$ and
$\beta$=$(\pi/4),(3\pi/4)$, $\epsilon_T$ exhibits four peaks
indicating the maximum amount of quantum correlation in the GHZ states
$1/\sqrt{2}\bigg( \pm |010101\rangle \pm |101010\rangle\bigg)$
describing one particle in each partition.  On the other hand, the
geometric measure is zero only for those values of the phases reducing
the linear superposition of the four states in Eq.~(\ref{phi}) to a
single term. For $\alpha=\pi/2$, $\epsilon_G$ shows a local minimum
but unlike $\epsilon_T$ it is not vanishing thus indicating a degree
of non-separability different from zero.  Its maximum values are found
when $\beta$=$(\pi/4),(3\pi/4)$ and $\alpha$=$(\pi/4),(3\pi/4)$,
namely when all the moduli of the coefficients of the linear
superposition of Eq.~(\ref{phi}) become equal and $|\Phi\rangle$
reduces to $1/2\bigg(|010101\rangle \pm |101010\rangle\pm
|111000\rangle \pm |000111\rangle \bigg)$.  The disagreement between
the qualitative behavior exhibited by the tripartite negativity and
the geometric measure for this specific case is representative of the
discrepancy between the two entanglement notions. In fact,
$\epsilon_G$ quantifies the amount of quantum correlations between the
modes regardless the number of particles in any subset, while
$\epsilon_T$ measures the entanglement of the particles in the modes.

\section{Entanglement dynamics in three-boson- and
  three-fermion-QWs}\label{Phys}
In the last years, theoretical and experimental investigations have
shown how the quantum statistics, due to exchange symmetry, may play a
key role in the appearance of quantum correlations between
non-interacting identical particles in various systems, ranging from
the propagation of photons~\cite{Hanb} to electron transport in
integer quantum Hall effect~\cite{Neder}. Among the different physical
phenomena examined, the continuous-time quantum walks (QWs) have
received great interest since they represent an ideal laboratory to
observe many-particle quantum mechanical behavior and to implement
future quantum technologies ~\cite{Rai,Bromberg,Peruzzo,Lah}.
Specifically, the emergence of non classical correlations in QWs of
two photons in an array of coupled waveguides has experimentally been
observed~\cite{Bromberg,Peruzzo}.  By using the criterion based on TPN
described in the previous section, here we investigate the
entanglement dynamics of the continuous-time QWs of three bosons and
of three fermions in an array of six sites.

In order to study the three-particle entanglement dynamics, we follow
the formalism developed in Ref.~\cite{Rai} for the QWs of
non-classical light in an array of coupled waveguides. We consider a
one-dimensional tight binding model, within the approximation of
nearest-neighbor interaction, described by the Hamiltonian
$\mathcal{H}$:
\begin{equation} \label{ham} \mathcal{H}=G\sum_{i=1}^6
  c_i^{\dag}c_i+T\sum_{i=1}^5(c_i^{\dag}c_{i+1}+c_{i+1}^{\dag}c_i),
\end{equation}
where $c_i^{\dag}$ $(c_i)$ is the creation (annihilation) operator for
a particle in the $i$-th site which obeys the commutation relation
$[c_i,c_j^{\dag}]_{\pm}$=$\delta_{ij}$, where the upper sign holds for
bosons and the lower one for fermions~\cite{Note1}. $G$ is the on-site
energy and $T$ is the tunneling rate at which the particles are
transferred to the neighboring sites. Here we are considering a
lattice with reflecting boundary conditions, so that the transition
amplitude from the node 1(6) to the node 6(1) is zero.  As it will be
shown in the following, this implies that the time evolution of the
quantum system is not periodic unlike the QWs of particles in lattices
with periodic boundary conditions~\cite{Mul,Xin}.

By diagonalizing the Hamiltonian~(\ref{ham}), in the Heisenberg
picture the annihilation operator $c_r$ at time $t$ can be obtained
as~\cite{Rai}:
\begin{eqnarray} \label{coeff}
  c_r(t)&=&\sum_{s=1}^6 c_s(0) C_{rs}, \nonumber \\
  C_{rs}&=&\frac{2}{7}\exp{\left(-\frac{iGt}{\hbar}\right)}\sum_{k=1}^6
  \exp{\left[-\frac{2itT} {\hbar}\cos{\left( \frac{k\pi}{7}\right)}
    \right]} \sin{\left( \frac{rk\pi}{7}\right)}\sin{\left(
      \frac{sk\pi}{7}\right)},
\end{eqnarray}
where $\frac{2}{7}\sum_{k=1}^6\sin{\left(
    \frac{rk\pi}{7}\right)}\sin{\left(
    \frac{sk\pi}{7}\right)}$=$\delta_{rs}$. Since any input Fock state
can be expressed by means of the creation operators $c_r^{\dag}$ and
the vacuum state $|0,0,0,0,0,0\rangle$, its time evolution during the
QWs can be calculated by using the Eq.~(\ref{coeff}). Even if the
effect of different initial occupation numbers on the entanglement
dynamics could be considered, here, for simplicity, we only examine
the case of three particles (bosons or fermions) initially coupled to
three neighboring sites, i.e. the state
\begin{equation}\label{stat}
  |\Psi(t=0)\rangle=|1,1,1,0,0,0\rangle.
\end{equation}

By means of the time evolution of the quantum state $|\Psi(t)\rangle$,
we can quantify the tripartite entanglement dynamics of the system by
using the TPN defined in Eq.~(\ref{enta3}). To this purpose, a
numerical approach has been used which allows one to evaluate at any
$t$ from $|\Psi(t)\rangle$ the density matrix $\rho_{1,1,1}$ where the
LPN of each subsets is fixed to 1 and then to diagonalize its partial
transposes for the evaluation of the bipartite negativities. Here we
consider two different partitions of the system, namely $A$=$\{1,2\}$,
$B$=$\{3,4\}$, $C$=$\{5,6\}$ and $A^{\prime}$=$\{1,4\}$,
$B^{\prime}$=$\{2,5\}$, $C^{\prime}$=$\{3,6\}$.

\begin{figure}[h]
  \begin{center}
    \includegraphics*[width=\linewidth]{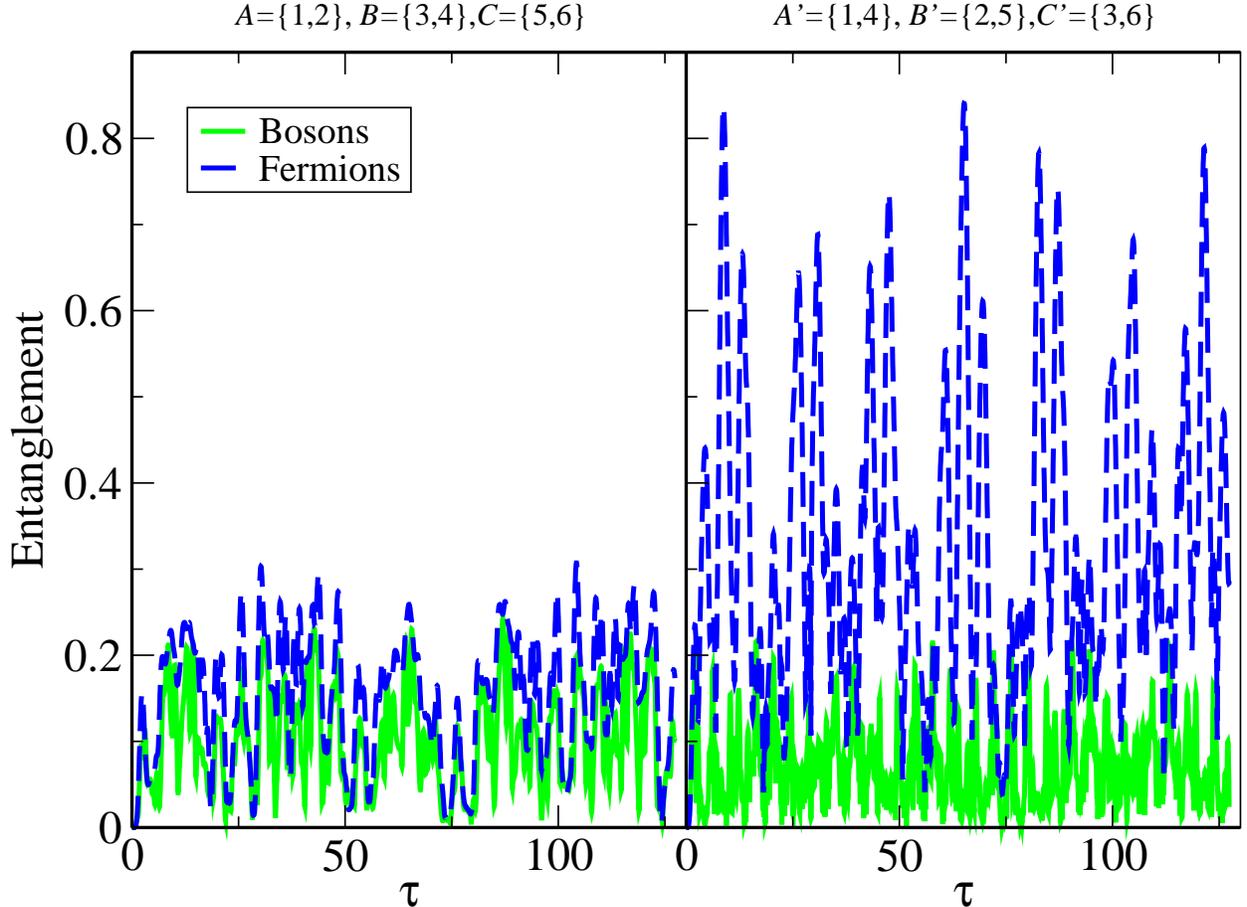}
    \caption{\label{fignew} (Color online) Entanglement as a function
      of the ``time'' $\tau$ for the fermionic (blue line) and bosonic
      (green line) systems when two different partitions are
      considered: $A$=$\{1,2\}$, $B$=$\{3,4\}$, $C$=$\{5,6\}$ (left
      panel) and $A^{\prime}$=$\{1,4\}$, $B^{\prime}$=$\{2,5\}$,
      $C^{\prime}$=$\{3,6\}$ (right panel). }
  \end{center}
\end{figure}

In Fig.~\ref{fignew} we report the entanglement of the three-boson and
the three-fermion state as a function of the ``time'' $\tau=tT/\hbar$
for the two above partitions.  We note that in all the cases the
amount of quantum correlations in the tripartite system under
investigation oscillates with time though such oscillations are not
periodic. In fact the interference of the three-particle wavefunction
amplitudes, due to the backscattering at reflecting boundaries,
results into a non-periodic time evolution of the system. Unlike the
case of single-particle dynamics in QWs into periodical
structures~\cite{Mul,Xin}, a time revival, namely the time interval
needed to reconstruct the whole initial wavefunction, cannot be found
here.

For both of the partitions, we note that the tripartite entanglement
for the fermions is always greater than the one exhibited by the
bosonic systems. This behavior can be explained by taking into account
the quantum statistics of the particles. The Bose-Einstein particles
exhibit bunching thus making finite the probability to find more
particles in the same site or in the same subset during the time
evolution of the system.  On the other hand, due to the exclusion
Pauli principle, no more than one fermion can occupy the same mode.
This implies that the term $P_{1,1,1}$ describing at any time the
probability to find one particle in each subset and appearing in the
definition of TPN of Eq.~(\ref{enta3}), increases moving from the
boson- to the fermion-system, even if both of them are initially
described by the same state $|\Psi(t=0)\rangle$. As a consequence, the
tripartite entanglement among bosons results smaller than the one
among fermions.

\begin{figure}
  \begin{center}
    \includegraphics*[width= \textwidth]{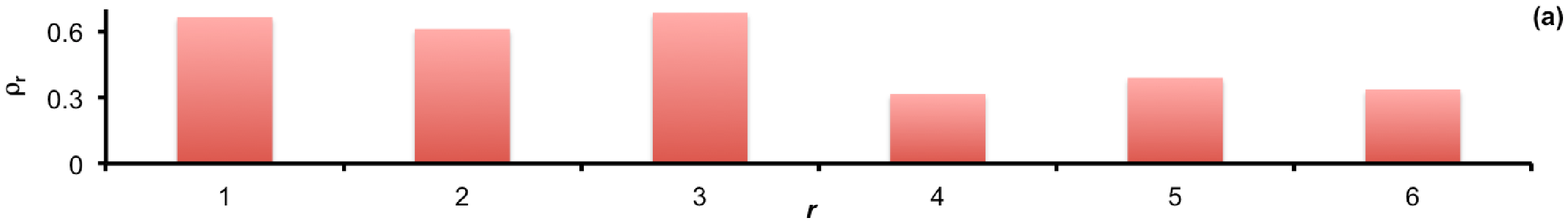}
    \begin{minipage}[c]{0.48\textwidth} 
      \includegraphics*[width= \textwidth]{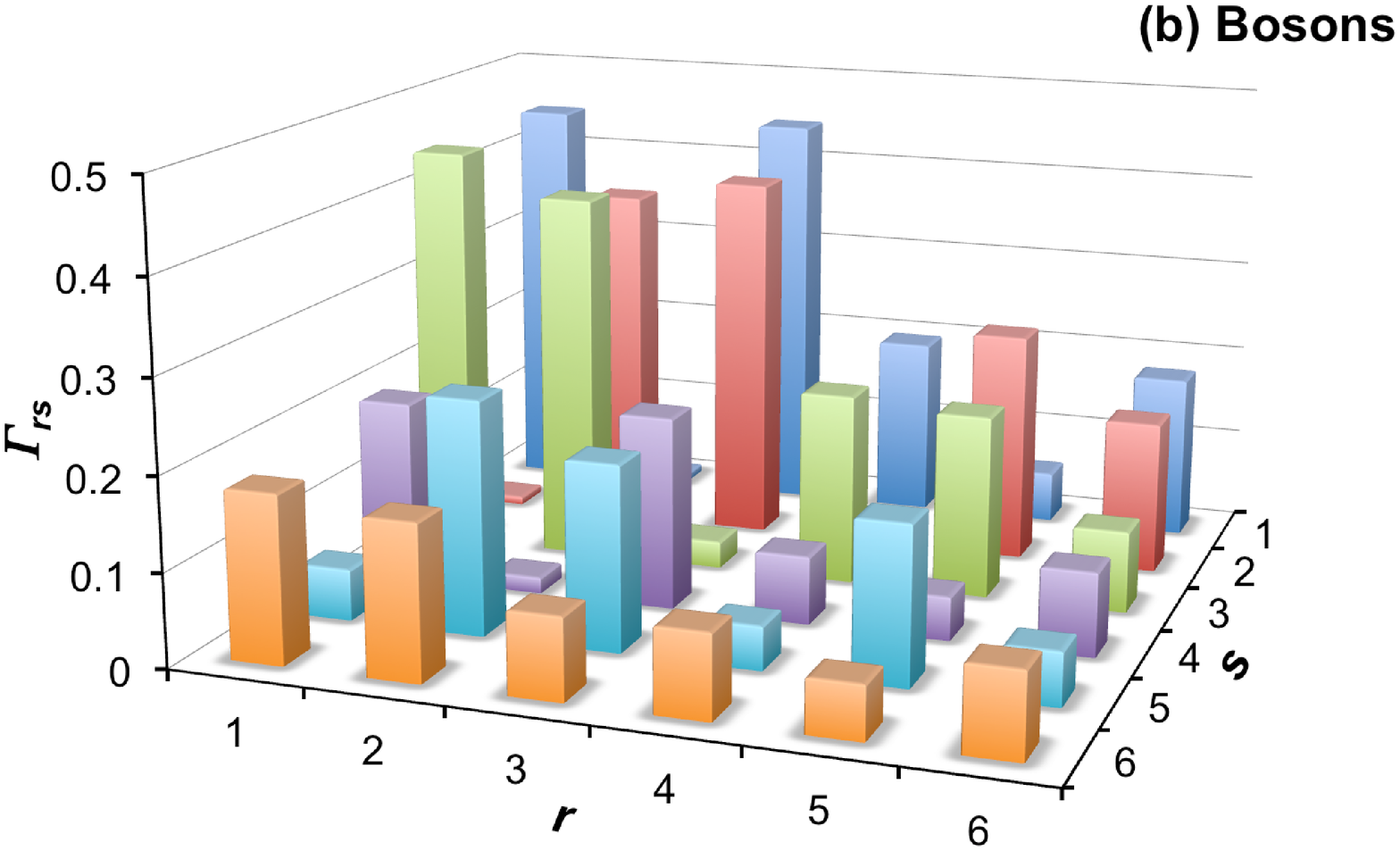}
    \end{minipage}
    \begin{minipage}[c]{0.48\textwidth}
      \includegraphics*[width=\textwidth]{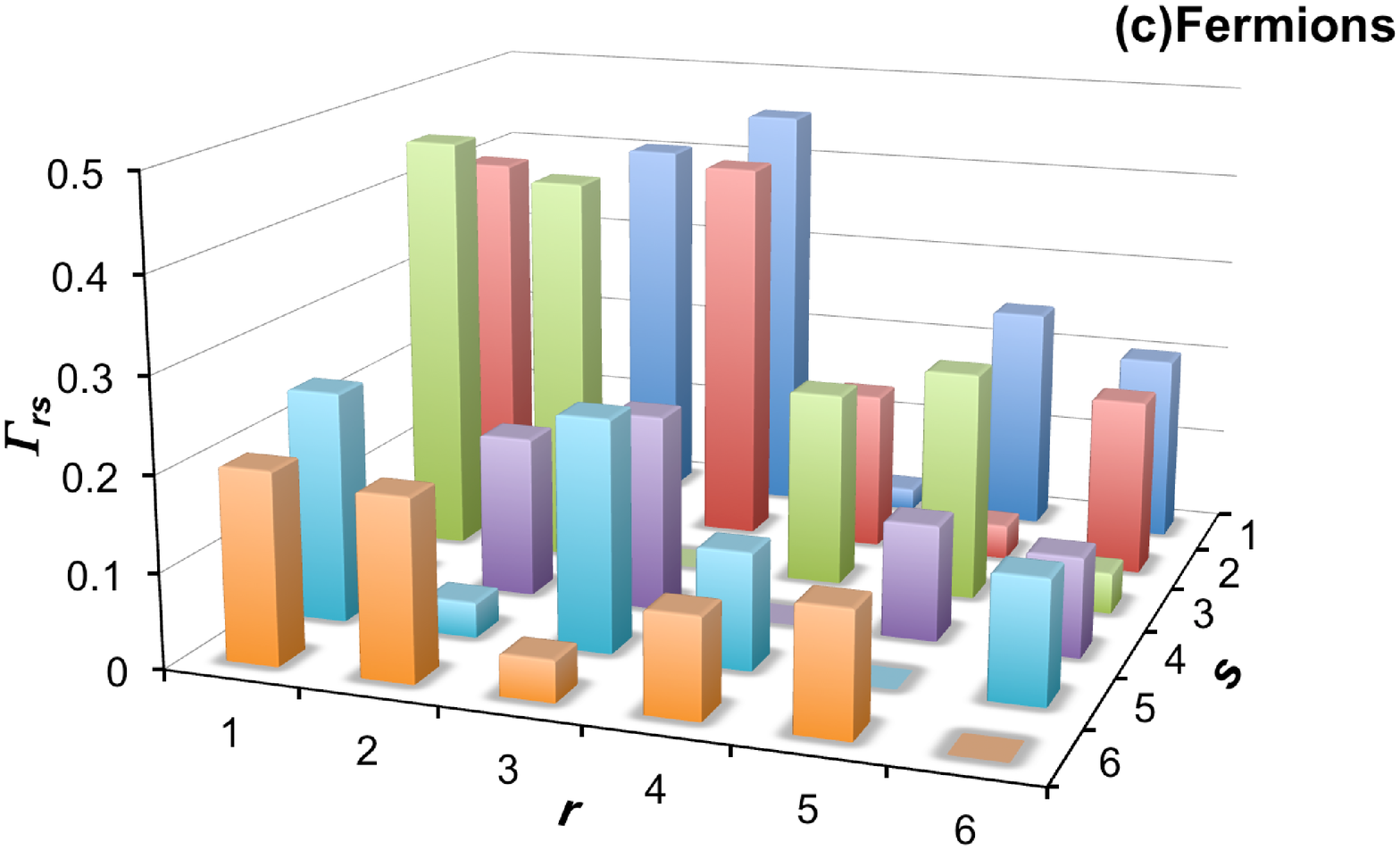}
    \end{minipage}
    \begin{minipage}[c]{0.48\textwidth} 
      \includegraphics*[width= \textwidth]{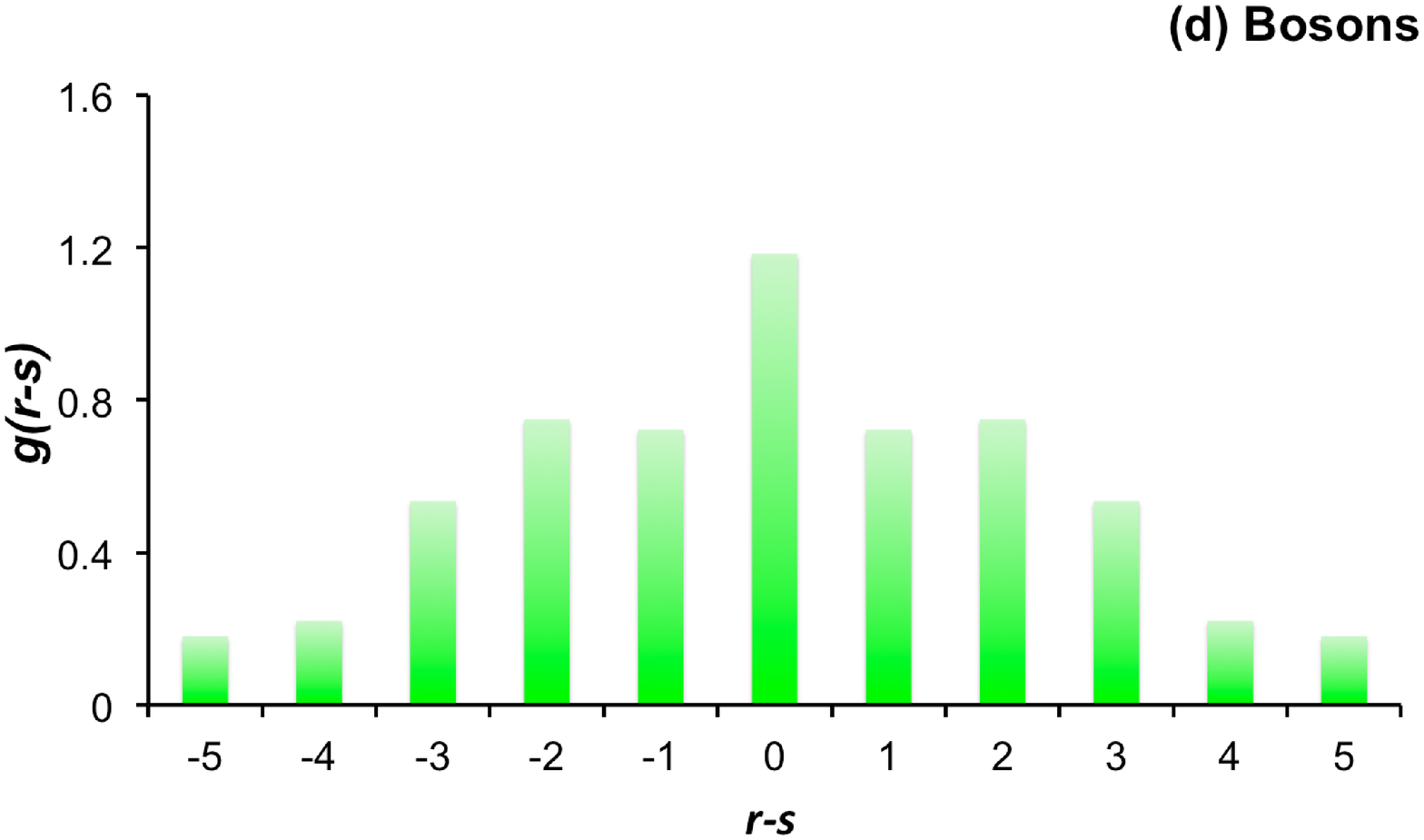}
    \end{minipage}
    \begin{minipage}[c]{0.48\textwidth}
      \includegraphics*[width=\textwidth]{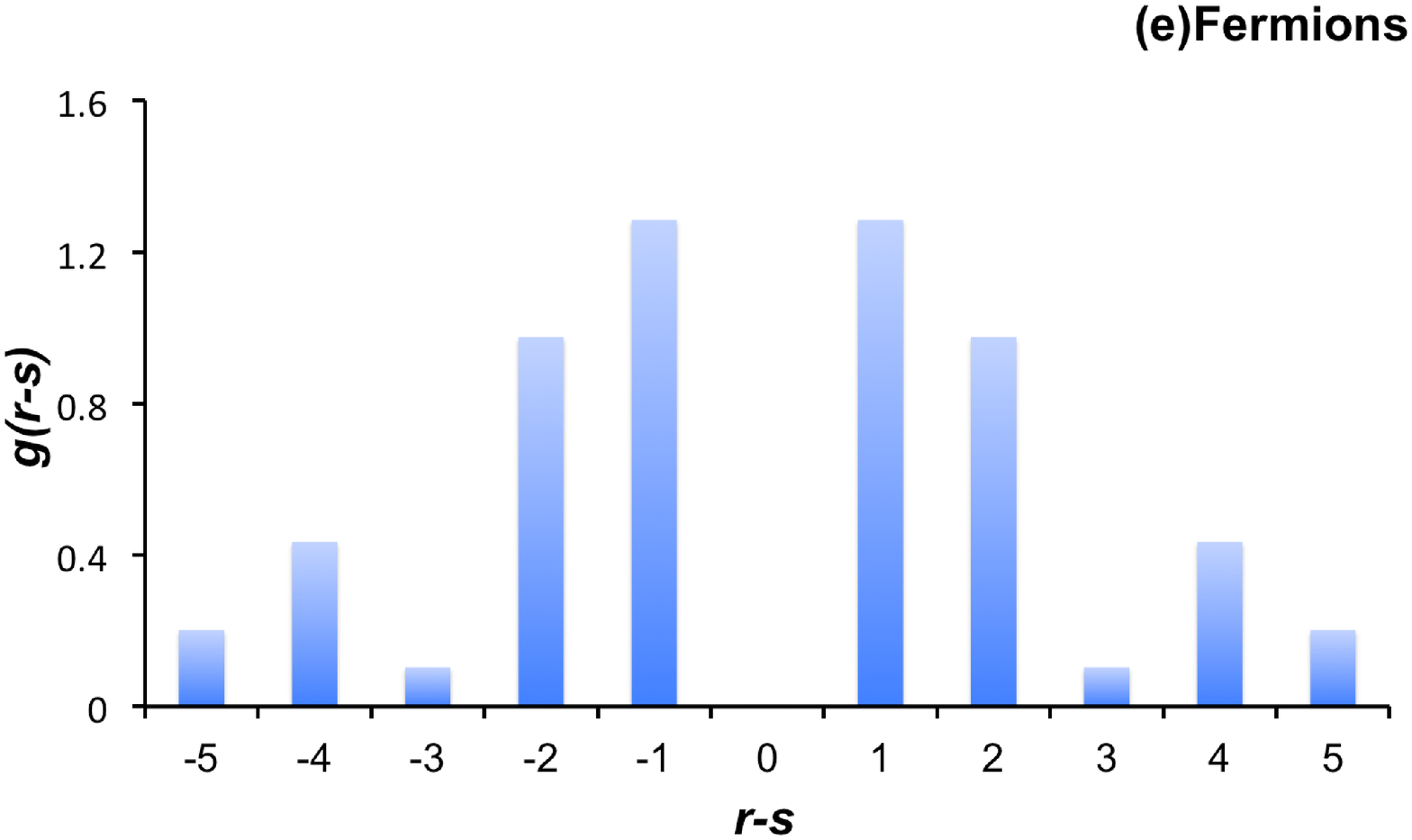}
    \end{minipage}
    \caption{\label{fig2} (Color online) Single and two-particle
      density probability of the bosonic and fermionic systems
      initially described by the state $|\Psi\rangle$ of
      Eq.~(\ref{stat}) for $\tau$=8.7. (a) The single particle density
      distribution $\rho_r$ which is identical for fermions and
      bosons.  (b) The correlation matrix $ \Gamma_{rs}$ for bosons
      computed from Eq.~(\ref{correb}). Bosons tend to localize in the
      the same or different sites of the first half of the lattice.
      (c) The correlation matrix $ \Gamma_{rs}$ for fermions computed
      from Eq.~(\ref{corref}).  Fermions are found with high
      probability in the neighboring sites of the first half of the
      lattice. (d) The interparticle distance probability for bosons.
      They tend to appear at same site or at neighboring sites. (e)
      The interparticle distance probability for fermions. They are
      more likely to be separated by a small number (1-2) of sites.}
  \end{center}
\end{figure}

The TPN dynamics of the two partitions examined presents some
differences.  While the tripartite bosonic entanglement shows the same
qualitative behavior, that is oscillations within the interval ranging
from 0 to 0.2, the amount of quantum correlations among fermions in
the partition where subsets are composed of non adjacent modes
results, on the average, larger than the one found for subsystems
controlling neighboring sites.  In order to deeper investigate this
behavior, we focus on some single- and two-particle features of the
quantum system at ``time'' $\tau=8.7$ where the fermionic TPN between
$A^{\prime}$, $B^{\prime}$, $C^{\prime}$ shows a peak. In
Fig.~\ref{fig2} the single particle density $\rho_r(t)=\langle
c_r^{\dag}c_r\rangle$ and the two-particle correlation
$\Gamma_{rs}(t)= \langle c_r^{\dag}c_s^{\dag}c_s c_r\rangle$ are
reported.  Here we do not consider the three-particle correlation
function, since the two-particle one is sufficient to understand the
features of the entanglement, as will be shown in the following.
$\rho_r(t)$ does not permit to analyze the quantum properties of the
system under investigation since it shows the typical features of the
classical QWs of two uncorrelated particles.  As a matter of fact, the
single-particle density whose expression in terms of the coefficients
$C_{rs}$ reads $\rho_r(t)=\sum_{s=1}^6 |C_{rs}|^2 n_s$ (with $n_s$ the
number of particles in the site $s$ at $t$=0) does not depend upon
quantum statistics and turns out to be identical for bosons and
fermions (see panel (a) of Fig.~\ref{fig2}).

On the other hand, in analogy with the Glauber photodetection
theory~\cite{Glauber} $\Gamma_{rs}(t)$ indicates the probability to
detect at time $t$ one particle at site $r$ and the other one at site
$s$ and displays the two-particle quantum correlations between the
locations $r$ and $s$ depending on the quantum statistics of the
particles.  For bosons it is given by
\begin{equation} \label{correb} \Gamma_{rs}(t)=\sum_{p=1}^6
  \sum_{q=1}^{p-1}\left|C_{rp}C_{sq}+C_{rq}C_{sp}\right|^2
  n_pn_q+\sum_{p=1}^6 |C_{rp}|^2|C_{sp}|^2 n_p(n_p-1),
\end{equation}
while for fermions it reduces to
\begin{equation}\label{corref}
  \Gamma_{rs}(t)=\sum_{p=1}^6 \sum_{q=1}^{p-1}\left|C_{rp}C_{sq}-C_{rq}C_{sp}\right|^2 n_pn_q.
\end{equation}
As shown by the correlation matrix in the panel (b) and (c) of
Fig.~\ref{fig2}, bosonic bunching results into a very large
probability to find two particles in the same sites in the first half
of the lattice. Instead if the particles follow the Fermi-Dirac
statistics, they tend to separate to neighboring sites of the first
half of the lattice (see panel (c) of Fig.~\ref{fig2}).  Regardless
the quantum statistics, this implies that for the partition
$\{A,B,C\}$ the probability that more particles occupy the same subset
is very high and, in turns, this means a low value of the entanglement
as depicted in the left panel of the Fig.~\ref{fignew}.  On the
contrary, for the partition where the subsystems control nonadjacent
modes, the LPN of $A^{\prime}$,$B^{\prime}$, and $C^{\prime}$ is more
likely to be equal to 1 and therefore the TPN can reach high values
(see the right panel of the Fig.~\ref{fignew}).

Also the interparticle distance probability~\cite{Lah}
$g(\Delta)=\sum_q \Gamma_{q,q+\Delta}$, depicted in the panels (d) and
(e) of Fig.~\ref{fig2} supports our interpretation. Apart from the
behavior in $\Delta$=0, related to the role played quantum statistics
in determining the probability to find two particles in the same site,
$g$ is high for low values of the intersite distance both for bosonic
and fermionic system thus confirming the localization of particles in
neighboring sites.

A more exhaustive discussion of the entanglement dynamics in our model
undoubtedly requires further analyses also on three-particle quantum
properties of the system at any time but this goes beyond the scopes
of this work.

\section{Conclusions}\label{Conclu}
In summary, we have presented a criterion for the tripartite
entanglement of mixed states of indistinguishable particles (both
bosons and fermions).  The entanglement measure proposed deviates from
the other approaches recently advanced~\cite{Lari}. They rely on the
entanglement of modes~\cite{Zana} but their physical significance is
not always very clear as shown here by some misleading results. As a
matter of fact, such measures can give values different from zero even
when applied to single-particle states which describe a particle in
equal superposition of modes. In this paper, we developed a different
criterion based on the notion of bipartite entanglement of particles
as introduced by Wiseman and Vaccaro~\cite{Wise}.  It provides a valid
guideline to quantify the amount of quantum correlations in tripartite
systems on which the allowed physical operations, such as preparation,
manipulation, measurement, do not change the local number of particles
contained in each subset.

Specifically, we have shown that a good entanglement measure
satisfying the requirements of the theoretical criterion introduced in
this paper can be obtained from the TPN recently used in the context
of non identical particles~\cite{Sabin,Anza}. Indeed, it constitutes a
very practical tool to evaluate the amount of quantum correlations
appearing in three-boson or -fermion mixture states of low-dimensional
systems, that can be extracted and then placed in standard quantum
registers without any violation of the local particle number
superselection rule.  As an application, we have considered a simple
toy-model of three fermions in a system with six modes and proved the
realibility of the TPN in comparison with the geometric multipartite
entanglement measure introduced in Ref.~\cite{Lari}.  In agreement
with theoretical predictions, we found that TPN is greater than the
geometric measure. The qualitative behavior of the two criteria as a
function of the parameters of the model is quite different thus
indicating their conceptual discrepancy. The geometric measure gives
an estimation of the amount of quantum correlations among the
occupation numbers of the sites regardless the particles.  On the
other hand the TPN, even if it depends upon the local particles number
in the subsystems, quantifies the entanglement among the particles
themselves.

Furthermore, we have used the TPN to quantify the time evolution of
the quantum correlations for the continuous-time QWs of three non
interacting particles (both bosons and fermions) in a six-mode
lattice, for two different partitions of the system. As expected, we
find that the entanglement dynamics is not periodic due to the
presence of reflecting boundary conditions in the lattice sites.
Furthermore, it strongly depends upon the quantum statistics of the
particle involved in the process: the bosonic bunching leads to a
large probability to find more bosons in the same site and therefore
in the same subset of the system. This implies a local particle number
equal to zero for another of the two parties and therefore the
separability of the three-particle state.  On the other hand, for
fermions the Pauli exclusion principle prevents occupation number of
sites greater than one, thus making more likely to find correlated
states with one particle in each subset.

Although the criterion introduced in this paper has been mainly
applied to the investigation of the tripartite entanglement in some
simple models, such as continuous-time QWs, it is completely general,
and its application turns out to be helpful to investigate the
building up of quantum correlations in both bosonic and fermionic
systems. In particular, it would be of interest to extend the TPN to
those physical system with large dimensionality where new and
interesting results could be expected due to higher complexity of the
systems themselves and to consequent richer structure of the quantum
correlations.

\begin{acknowledgments}
  The authors would like to thank A.~Bertoni, and C.~Jacoboni for
  fruitful discussions.
\end{acknowledgments}

\end{document}